\documentstyle[preprint,aps]{revtex}

\begin{document}

\thispagestyle{empty}
\setcounter{page}{1}

\title{ DISPERSION RELATIONS
IN ULTRADEGENERATE RELATIVISTIC  PLASMAS} 

\author{Cristina Manuel\footnote{E-mail: Cristina.Manuel@cern.ch} }

\address{Theory Division, CERN, CH-1211 Geneva 23, Switzerland}

\maketitle

\thispagestyle{empty}
\setcounter{page}{0}

\begin{abstract}
$\!\!$ The propagation of excitation modes in a relativistic
ultradegenerate plasma is modified by their interactions with
the medium. These modifications can be computed by evaluating
their  on-shell self-energy, which gives
(gauge-independent) dispersion relations. For modes with momentum
close to the Fermi momentum, the one-loop fermion self-energy is dominated
by a diagram with a soft photon in the loop.
We find the one-loop dispersion relations for quasiparticles and
antiquasiparticles, which behave differently as a consequence
of their very different phase-space restrictions when they
scatter with the electrons of the Fermi sea. In a relativistic
system, the unscreened magnetic interactions spoil the normal
Fermi liquid behavior of the plasma. For small values
of the Fermi velocity, we recover the non-relativistic dispersion
relations of condensed matter systems.
\end{abstract}

\vfill

\hfill\break
\hbox to \hsize{CERN-TH/2000-132} 
\hbox to \hsize{May/2000}
\vskip-12pt
\eject

\baselineskip= 20pt
\pagestyle{plain}
\baselineskip= 20pt
\pagestyle{plain}

\section{Introduction}

The study of hot relativistic plasmas is  nowadays a very active field of research
\cite{LeBellac}. This is due to the existence of experimental programs to test the
existence of the quark-gluon plasma phase of QCD. The physics of some
astrophysical settings, such as those of neutron stars and supernovas, 
also requires knowledge  of a different regime  of relativistic plasmas,
less hot but still very dense.
This cold and ultradegenerate regime of QED and QCD
has been much less explored. However, the fact that
matter at  very high baryonic densities behaves as a color superconductor has 
given us a strong motivation to study ultradegenerate
relativistic plasmas, as  several new phenomena occurs in this phase of
QCD (see \cite{Wilczek:1999ym} and \cite{Rischke:2000pv}, and references
therein).

One of the central concepts in a plasma is that of a quasiparticle.
A particle immersed in a medium modifies its propagation
 properties by interacting with the surrounding medium.
In  field theoretical language, we would say
that the particle is ``dressed" by a self-energy cloud. 
In the ultradegenerate plasma the  relevant degrees of freedom 
are those of quasiparticles or quasiholes
(absences of particles in the Fermi sea) living close to the Fermi surface.
Because of the exclusion principle, quasiparticles/quasiholes can only live if 
they are outside/inside the Fermi sea, respectively.
These excitations tend to lower their
energy, by  undergoing collisions with the particles in the Fermi
sea. They  decay, and  thus have a finite lifetime.
The concept of quasiparticle,
however, only makes sense if its lifetime is long enough, or in other words, if its
damping rate is much smaller than its energy.

Here we will mainly be concerned with electromagnetic plasmas. 
There is a vast literature on the quasiparticle properties in 
non-relativistic cold plasmas \cite{Pines}.
The same does not hold true for the relativistic ones, though.
There are two main differences in these two  energy regimes
of a plasma. In the non-relativistic domain, the
electric interactions are dominant, while the magnetic ones are
suppressed by a factor $(v/c)^2$, where $v$ is the velocity of the particle, 
and $c$ is the velocity of light. Thus, magnetic interactions start to
be relevant only when quasiparticles are  fast enough or, in other words,
when the Fermi velocity $v_F$ approaches the velocity of light. 
This is an important difference, as 
electric interactions are not long-ranged in the medium, because of Debye
screening, while magnetic interactions are. The relevance of this
last point has already been stressed in the condensed matter literature
\cite{Holstein}, just noticing that magnetic interactions spoil the
normal Fermi liquid behavior of the plasma. The second main difference
is due to the fact that in a relativistic plasma there are also antiparticle
excitations. Their contribution to any physical process
is in general suppressed,
since it takes  more energy to excite an antiparticle
than a particle of the Fermi-Dirac sea. Nevertheless, in the context
of the color flavor locking phase of QCD \cite{Alford:1999mk},
some of the properties
of the antiparticles determine the mass spectrum of the Goldstone modes
which arise from the spontaneous breaking of chiral symmetry, so
those cannot be neglected.

In this paper we study the quasiparticle and antiquasiparticle
 dispersion relations in a
full relativistic framework, generalizing the results of a previous
publication \cite{LeBellac:1997kr} to the case where the Fermi velocity
$v_F \neq c$. We  can thus explore all the energy domains of the system,
and in particular,  we can take the non-relativistic
limit $v_F \ll c$, and match the results obtained in the
condensed matter literature
\cite{Pines,Holstein}.
The dispersion relations are obtained by computing the on-shell one-loop
self-energy. While the one-loop self-energy
is in general gauge-dependent, 
it is not when evaluated on the particles mass-shell. 
For quasiparticles with momenta close to the Fermi momentum, the one-loop
self-energy is dominated by a diagram in which the photon is soft. 
When the photon is soft, it also needs  to be dressed to take properly into
account  the effects of the medium. This can be done by
using the resummation techniques proposed by Braaten and Pisarski
\cite{Pisarski:1989vd}, and considering hard thermal loop 
photon propagators, or hard dense loop (HDL) ones for the ultradegenerate
case \cite{Manuel:1996td}. We first compute the on-shell
imaginary part of the one-loop self-energy for electrons and positrons,
which can be interpreted in terms of their scattering  with particles
of the Fermi sea, via an exchange of a soft photon. The on-shell
real part of the self-energy can be reconstructed from the on-shell
imaginary part, just by using a Kramers-Kroning relation.
 
This paper is structured as follows. Section II introduces the notation
of the paper. We work in natural units, $c=\hbar=k_B =1$, unless otherwise
stated.
 In Sect. IIIA we compute the on-shell one-loop self-energy
of the fermion. We take the non-relativistic limit of our results in
Sect. IIIB, and conclude in Section IV. In  Appendix A the spectral functions of the HDL photon propagators are given for $v_F \neq c$, and in
 Appendix B Luttinger's theorem is recalled.

\section{Dispersion Relations for the quasiparticles}

We consider a plasma with a finite density of electrons, characterized
by a chemical potential $\mu$. In order to guarantee its stability,
we assume that the electrons are immersed in a uniform background of 
positive charges, of density equal to the average electron density. 
These background charges can be due to positively charged ions, which
are very heavy.

In a  plasma with chemical potential $\mu$, the propagation properties of the 
quasiparticles are modified by medium effects. The dressed fermion propagator
$S(P)$, where $P=(p_0, {\bf p})$ is the four momentum,
obeys the Schwinger-Dyson
equation
\begin{equation}
S^{-1}(P) = S_0 ^{-1}(P) + \Sigma(P) \ ,
\label{eq1}
\end{equation}
where $S_0^{-1}$ is the inverse  free propagator
\begin{equation}
\label{eq2}
S_0 ^{-1}(P) = {P\llap{/\kern1pt}} + \mu \gamma_0 - m \ ,
\end{equation}
with ${P\llap{/\kern1pt}} = P^\mu \gamma_\mu$ 
and $\Sigma (P)$ is the one-loop self-energy. 

Because of the clear asymmetry between electrons and positrons in the
electromagnetic plasma,
it is convenient to treat them separately, as their propagation properties will
be modified in different ways.
Introducing  the positive and negative energy projectors
\begin{equation}
\label{eq3}
\Lambda_{\bf p} ^{\pm} = \frac{E_p    \pm \left( \gamma_ 0 {\bf \gamma} \cdot
{\bf p} + m \gamma_0 \right)}{2 E_p} \ ,
\end{equation}
where  $E_p = \sqrt{ p^2 + m^2}$,  we can rewrite   
\begin{eqnarray}
S_0 ^{-1}(P) & = & \gamma_0 \Lambda_{\bf p}^+ \left(p_0 +\mu - E_p \right) +
\gamma_0 \Lambda_{\bf p} ^- \left(p_0 + \mu + E_p \right) \ ,  \\
\Sigma(P) & = & \gamma_0 \Lambda_{\bf p}^+ \Sigma_+ (P) -
\gamma_0 \Lambda_{\bf p}^-  \Sigma_-(P) \ .
\end{eqnarray}
After inverting  (\ref{eq1}) one gets
\begin{equation}
\label{eq6}
S(P) = S_+(P) \Lambda_{\bf p}^+ \gamma_0 + S_-(P) \Lambda_{\bf p}^- \gamma_0 \ , 
\end{equation}
where
\begin{equation}
\label{eq7}
S_{\pm} (P) = \frac{1}{p_0 + \mu \mp \left( E_p - \Sigma_\pm (P)\right)} \ ,
\end{equation}
and the upper/lower subscripts refer to electrons/positrons, respectively.

Every energy eigenstate can be projected onto states of given helicity,
with the projectors
\begin{equation}
{\cal P}^{\pm}( {\bf p}) = \frac{1 \pm \gamma_5 \gamma_0 {\bf \gamma \cdot {\hat p}}}{2} \ . 
\label{neweq}
\end{equation}
In principle, the most general structure of the one-loop self-energy
contains four unknown functions, according to the energies and helicities
of the quasiparticles. However, the effects which will
be discussed in this article do not depend on the helicity of the
quasiparticles, and thus we would not explicitly take into account the
helicity projectors.

The value of the one-loop self-energy $\Sigma$ is gauge-dependent. 
However, when it is evaluated on the particles mass-shell,
it should be gauge independent. This is so because the 
poles of (\ref{eq7}) give the physical dispersion relations of electrons and positrons 
which define their propagation properties in the plasma.

The dispersion relations obtained from (\ref{eq7}) 
are 
\begin{eqnarray}
\label{eq8}
\omega_\pm &=& -\mu \pm
\left(E_p  - {\rm Re}\,\Sigma_\pm(\omega_\pm + i \gamma_\pm, {\bf p}) \right) \ ,   \\
\gamma_\pm &=& \mp  {\rm Im} \Sigma_\pm(\omega_\pm + i \gamma_\pm, {\bf p}) \ ,
\label{eq9}
\end{eqnarray}
where $\omega_{\pm}$ and $\gamma_{\pm}$ define the energy and damping rates of the
electrons/positrons,  respectively. For the concept of quasiparticle to make sense,
it is necessary that $\gamma_{\pm} \ll \omega_{\pm}$, so that the quasiparticles are long-lived enough.

In the remaining part of the paper the dispersion relations
for quasiparticles and antiquasiparticles
with  momentum close to the Fermi momentum
will be studied. In this case,
the dominant contribution to their one-loop self-energy arises when the
photon in the loop is soft, that is, of order $\sim e \mu$, where $e$ is the electromagnetic coupling constant. 
When the photon is soft it has also to be dressed, in order to take into account
properly the medium effects of Debye screening and Landau damping.

\section{The on-shell fermion Self-Energy }

\subsection{Relativistic Domain}

For a plasma at temperature $T$ and chemical potential
$\mu$, we compute  the one-loop self-energy $\Sigma$
using the imaginary time formalism.
It is  convenient to use the spectral function representation of the
 fermion and photon propagators in the computation.
 The free fermion propagator is given by
\begin{equation}
S_0(i\omega_n,{\bf k})=\int_{-\infty}^\infty{{\rm d}k_0\over
2\pi}\,{( {K\llap{/\kern3pt}}+m)\rho_f(K)\over
k_0-i\omega_n-\mu} \ , 
\label{eq10}
\end{equation}
with
\begin{equation}
\rho_f(K)=\frac{\pi}{E_k} \left(\delta(k_0-E_k) -\delta(k_0+E_k) \right)  \ .
\label{eq11}
\end{equation}
In (\ref{eq10}), $\omega_n=\pi(2 n+1)T$ is a fermionic Matsubara frequency. 
The (resummed) photon propagator $\Delta_{\mu\nu}(Q)$,
where $Q=(i \omega_s, {\bf q})$,  and $\omega_s = 2 \pi s T$ is a bosonic Matsubara frequency,
 is written in the Coulomb gauge
\begin{equation}
\Delta_{\mu\nu}(Q)=\delta_{\mu 0}\delta_{\nu 0} \, \Delta_L(Q)
+{\cal P}^T_{\mu \nu}\Delta_T(Q) + \xi_C \frac{Q_\mu Q_\nu}{q^4}\ ,
\label{eq12}
\end{equation}
where  ${\cal P}^T _{i j} = (\delta_{ij}-\hat q_i\hat q_j)$,
 ${\hat q}^i = {\bf q}^i/|{\bf q}|$, 
  ${\cal P}^T_{i0} = {\cal P}^T_{0i} =
{\cal P}^T_{00}=0$, and $\xi_C$ is the gauge parameter. 
The longitudinal and transverse propagators are written in terms of their spectral functions
\begin{mathletters}
\label{eq13}
\begin{eqnarray}
\Delta_L (i \omega_s, q)   & = & 
\int_{-\infty}^{\infty}{{{\rm d} q_0 \over 2\pi}  \frac{ \rho_L (q_0,q) } {q_0-i \omega_s }}  
- {1 \over q^2 } \ ,  \\
 \Delta_T  (i \omega_s, q) &  =  &
\int_{-\infty}^{\infty}{{{\rm d} q_0 \over 2\pi}  \frac{ \rho_T (q_0,q) } {q_0-i \omega_s}} 
 \ .
\end{eqnarray}
\end{mathletters}
$\!\!$Analytical  expressions for  $\rho_{L,T}$ can be found in
\cite{Pisarski:1989cs,LeBellac} for the case of an ultrarelativistic ($m=0$) plasma.
At $T =0$, it is also possible to derive the
spectral functions for $m \neq 0$ \cite{Manuel:1996td}.
We present analytical expressions
for the spectral functions in this case  in  Appendix A.

The one-loop self-energy  
\begin{equation}
\Sigma(P)= e^2T\sum_s\int{{\rm d}^3q\over (2\pi)^3}\gamma_\mu 
S_0(P-Q)\gamma_\nu 
\Delta_{\mu\nu}(Q) \ ,
\label{eq14}
\end{equation}
when expressed in terms of the spectral functions, reads
\begin{eqnarray}
\label{eq15}
\Sigma(i \omega, p) &=& e^2T\sum_n\int{{\rm d}^3q\over (2\pi)^3} \int^{\infty}_{-\infty}
\frac{{\rm d} k_0}{2 \pi} \rho_f(K) \,\gamma_\mu  ( {K\llap{/\kern3pt}}+m)\gamma_\nu \\
& \times & \left\{ \left( \int^{\infty}_{-\infty}
\frac{ {\rm d}q_0}{2 \pi} \frac{ \delta_{\mu 0} \delta_{\nu 0}
\rho_L(q_0,q) + {\cal P}^T_{\mu \nu} \rho_T(q_0,q)
}{\left(q_0 -i \omega_n \right)
 \left(k_0 - i \omega + i \omega_n - \mu  \right)}  \right)
- \frac{1}{q^2} \frac{\delta_{\mu 0} \delta_{\nu 0}}
{\left(k_0 - i \omega + i \omega_n - \mu  \right)} \right.  \nonumber \\
 &+& \left.\xi_C  \frac{Q_\mu Q_\nu}{ q^4} \right\} \nonumber \ .
\end{eqnarray}
The sum over Matsubara frequencies is now easily performed. After 
 analytical continuation $i\omega_n+\mu\to
p_0+i\eta$, with $\eta \rightarrow 0^+$ to Minkowski space, one can evaluate the
on-shell  imaginary part. It is very easy to realize that none of the 
last two pieces of Eq. (\ref{eq15}) contribute to this on-shell  imaginary part.
Therefore, the result of the computation is gauge independent. One finds
\begin{eqnarray}
\label{eq16}
{\rm Im}\,\Sigma(p_0 + i\eta, p) &=& -e^2 {\rm Im}\, 
 \int{{\rm d}^3q\over (2\pi)^3} \int^{\infty}_{-\infty}
\frac{{\rm d} k_0}{2 \pi} \rho_f(K) \int^{\infty}_{-\infty}
\frac{{\rm d} q_0}{2 \pi} 
\frac{1 + f(q_0) - {\tilde f}(k_0 -\mu)}{p_0 -k_0 -q_0 + i \eta} \\  
& \times &\gamma_\mu  ({K\llap{/\kern3pt}}+m) \gamma_\nu \left[\delta_{\mu 0}
 \delta_{\nu 0}
\rho_L(q_0,q) + {\cal P}^T_{\mu \nu} \rho_T(q_0,q)\right] \nonumber \ .
\end{eqnarray}

In (\ref{eq16}), $f$ and ${\tilde f}$ are Bose-Einstein and Fermi-Dirac
distribution functions ($\beta=1/T$)
\begin{equation}
f(q_0)={1\over {\rm e}^{\beta q_0}-1} \ , \qquad
{\tilde f} (k_0 -\mu)={1\over {\rm e}^{\beta (k_0-\mu)}+1} \ .
\label{eq17}
\end{equation}

The damping rates for the quasiparticles and antiquasiparticles
 are thus obtained
after multiplying
Eq.  (\ref{eq16}) by the corresponding projectors and taking a Dirac trace, evaluating
the final expression on the particles mass-shell\footnote{At this point, one
can check that the damping rate of the quasiparticles does not depend on
their helicities, by using the helicity projectors of Eq.(\ref{neweq}). }

\begin{equation}
\gamma_{\pm}  = \mp \,{\rm Im}\,\Sigma_\pm(p_0+i\eta\ ,{\bf
p}) \Big|_{p_0 \, {\rm on -shell}}   \ .
\label{eq18}
\end{equation}
To obtain the damping rate of a quasiparticle 
 one has to evaluate the imaginary part of its
self-energy on the  pole of the dressed  propagator. However,  
up to corrections of order $e^2$, it would be enough to consider the above expressions at $p_0 = \pm E$, as the corrections introduced by $\Sigma_{\pm}$ only
displace these poles by an amount proportional to $e^2$.
 In this case, after evaluating the Dirac traces we find,
with ${\bf k} = {\bf p} - {\bf q}$,
\begin{eqnarray}
\label{eq19}
\gamma_{\pm} & =&\pm {\pi e^2\over E} \int{{\rm d}^3q\over
(2\pi)^3}\int_{-\infty}^\infty{{\rm d}k_0\over 2\pi}\rho_f(k_0)
\int_{-\infty}^\infty{{\rm d}q_0\over 2\pi} \\
&\times &
\left(1+ f(q_0)- {\tilde f}(k_0-\mu)\right)  \delta(p_0-k_0-q_0)
\left\{[p_0k_0+{\bf p\cdot
k}+m^2] \right. \nonumber \\
&\times & \left. \rho_L(q_0,q)
+2[p_0k_0-({\bf p\cdot \hat q})({\bf k\cdot\hat
q})-m^2]\rho_T(q_0,q)\right\} \Big|_{p_0= \pm E}  \ .
\nonumber
\end{eqnarray}

Equation (\ref{eq19}) gives the general expression for the damping rates
for any value of $T$, $\mu$ and $m$.  At very high temperature, 
these damping rates  are infrared (IR) logarithmic
divergent, even after including the screening corrections.
This is due to the soft photon contribution,
as $f(q_0) \sim T/q_0$ for $q_0 \ll T$.
Perturbation theory fails to provide the damping rates at high $T$
\cite{Pisarski:1993rf}.
A non-perturbative treatment to resum the leading order divergences was proposed in
\cite{Blaizot:1996hd} to find a non-exponential decay law for the quasiparticles.

In this paper we are concerned with the ultradegenerate limit, when
$T=0$. In this case
several simplifications occur. The  damping rates are IR finite after
the inclusion of the screening effects \cite{LeBellac:1997kr},
 as opposed to what happens at high $T$.
For $T=0$, $(1+f(q_0))=\Theta(q_0)$, where $\Theta$ is the step function.
For $T=0$ the fermion distribution function is
${\tilde  f}(E_k-\mu)=\Theta(\mu-E_k)$, while
${\tilde  f}(-E_k-\mu)=1-{\tilde  f}(E_k+\mu)= 1 $. 

From this point on, it is convenient to treat separately the electron and
positron damping rates, as different phase-space restrictions arise in the two cases.
If we concentrate in the soft photon region, we can approximate
\begin{equation}
E_k = \sqrt{|{\bf p}-{\bf q}|^2 + m^2 } \simeq E - {\bf v} \cdot {\bf q} \ ,
\end{equation}
where ${\bf v} = {\bf p}/E$ is the velocity of the fermion.
We thus find
\begin{mathletters}  
\label{eq21}
\begin{eqnarray}
\gamma_{+} & \simeq & {e^2\over 8 \pi^2 v} \int_{ q \,{\rm soft}} 
q {\rm d}q\, {\rm d} q_0\left[   \left(\Theta(q_0) \, -\Theta(\mu-E+q_0) \right)  
  \left\{ \rho_L(q_0, q)+ v^2(1-\cos^2\theta)\rho_T(q_0,q) \right\} \right]
 \ ,  
\label{eq21.a} \\
\gamma_{-} & \simeq & - {e^2\over8 \pi^2 v } \int_{q \, {\rm  soft}}
q {\rm d}q \,{\rm d} q_0 \left[
  \Theta(-q_0)     
  \left\{ \rho_L(-q_0, q)+ v^2(1-\cos^2\theta)\rho_T(-q_0,q) \right\}\right]
 \ , 
\label{eq21.b}
\end{eqnarray}
\end{mathletters}
$\!\!$where $q_0 = q v \cos\theta$.

The damping rates for the electron and the positron thus only differ in the
 phase-space
 restrictions of these two types of particles. One can interpret the
above equations as follows. A particle/antiparticle, with energy 
$\pm E$ is scattered to a state of
energy $\pm E_k$, respectively,
 creating a particle-hole pair. For the electron, $E_k$ is forced to be
above the Fermi energy, because of Pauli blocking. This last restriction is absent in the case of the positron.

For a quasiparticle with velocity close to the Fermi velocity, we can further approximate $v \approx  v_F$ in Eq. (\ref{eq21.a}).
 From the fact that the spectral functions 
$\rho_{L,T}$ in Eqs. (\ref{eq21}) are evaluated for values of
 $q_0^2 \leq q^2 v^2_F$, we see
that it is only the part of the spectral function corresponding to Landau damping (the functions $\beta_{L,T}$ in 
(\ref{betas})), that contributes to the integrals. 
Using the explicit values of spectral densities as given in
 Appendix A, one can
evaluate the above integrals numerically. Analytical expressions can be obtained for
the interesting case  $|E-\mu| \ll M$ (so this includes the case of
quasiholes). In this
regime we find at leading order
\begin{equation}
\label{eq22}
\gamma_{ +} \sim \frac{e^2}{24 \pi} |E - \mu| +  \frac{e^2}{64 v^2_F  M } (E - \mu)^2 + {\cal O}(|E-\mu|^3) \ , 
\end{equation}
which generalizes the expressions obtained in Refs.
\cite{LeBellac:1997kr,Vanderheyden:1997bw}
 for the case
$v_F \neq 1$. In the above equations $M= \sqrt{e^2 \mu^2 v_F/ \pi^2}$
is the Debye mass. 
The first terms in the r.h.s. of Eqs. (\ref{eq22}) are due to
scattering processes with
exchange of soft magnetic
photons, while the second is due to the exchange of soft
 electric ones. As  can be seen
the magnetic contribution is suppressed with respect to the electric one by a factor
$v^2_F$. Therefore, the electric contribution is dominant for
$v_F \ll 1$.
In the ultrarelativistic limit, $v_F =1$, the damping rate of the electron
is dominated by the magnetic contribution.

The damping rate of a quasiparticle, or a quasihole, which lives close
to the Fermi surface can then be expressed as a power series in
$|E-\mu|$. If this parameter is large, then $\gamma_+$ is large, and
the lifetime of these excitations is so short, that it does not make
sense to talk about quasiparticles or quasiholes. On the contrary,
when the fermion energy approaches the Fermi energy, its lifetime tends
to infinity. In particular, from Eq. (\ref{eq22}) one deduces that 
the Fermi sea is stable. 
We should point out also the very different contribution to the
energy dependence of $\gamma_+$ from the electric and magnetic interactions.
The quadratic dependence on $(E - \mu)$
of $\gamma_+$ can be entirely understood as arising
from the short-ranged character of the electric interactions in the plasma,
and also the phase-space restrictions of electron-electron scattering 
(see Appendix B). Magnetic interactions are not short-ranged,
but  only suffer a  weak dynamical screening due to Landau damping.
The linear dependence on
$(E-\mu)$ is also a product of Landau damping and phase-space restrictions. 

We now consider the damping rate of the antiquasiparticle.
Let us first stress that for a positron pair annihilation also contributes to
its damping rate. However, this is a process that occurs at order $e^4$, 
and it can be computed by taking the imaginary part of a two-loop correction
to the fermion self-energy. In a weak coupling expansion, the damping
rate of the positron is dominated by the scattering of the positron with
the electrons of the Fermi sea. For a positron with velocity
 $v \approx  v_F$, we can evaluate numerically Eq. (\ref{eq21.b}). 
The only difference with respect to the computation of $\gamma_+$ comes
from the different phase-space restrictions for antifermions, or in
other words, the different domain of integration of the integrals. 
We find at leading order
\begin{equation}
\gamma_{-} \sim e^2 M \left(\frac{v^2_F}{24 \pi} + \frac{1}{64} \right) \ ,
\end{equation}
which agrees for $v_F=1$ with the result of Ref.\cite{Vanderheyden:1997bw}. 

The on-shell real part of the self-energy can be obtained from the
general expression (\ref{eq15}), just by using a principal value 
prescription to evaluate the integral after the analytical continuation
to Minkowski space is done. However, it is much simpler to reconstruct 
it from the value of the on-shell imaginary part, using a Kramers-Kroning
dispersion relation, which gives the value of the real part, up to a
constant.

If $f_{\pm}(\omega)$ is an analytic function in the upper/lower complex plane,
respectively, then
from the Cauchy theorem, its  real part is given as a  function of 
its imaginary part as
\begin{equation}
{\rm Re} f_{\pm}(\omega) = \pm \frac{ PP}{\pi}
 \int^{\infty}_{-\infty} \frac {{\rm d} \omega'}{ \omega' - \omega}
\, {\rm Im} f_{\pm}(\omega')  
+ C_\infty \ ,
\end{equation}
where $PP$ denotes the principal value of the integral along the real axis, and $C_\infty$ is
a subtraction constant needed in case that $f_{\pm}$ does not vanish for $|\omega| \rightarrow \infty$.

Since we have computed the damping rate $\gamma_+$ for values
$|E-\mu| \ll M$, the only energy domain where the concept of quasiparticle
makes sense, we will use the dispersion relation using a cutoff which
implements this constraint, that is with cutoffs
$\Lambda_{\pm} = \mu \pm M$. We thus find
\begin{equation}
\label{eq24}
{\rm Re} \Sigma_+(E,p) \sim {\rm Re} \Sigma_+(\mu,p) +
 \frac{e^2}{12 \pi^2} (E-\mu) \ln {\frac{
M}{|E-\mu|}} + \frac{e^2}{32 \pi v^2_F} |E-\mu| + 
{\cal O}((E-\mu)^2) \ .
\end{equation}
The value of the energy-independent constant ${\rm Re} \Sigma_+(E=\mu,p)$, 
which renormalizes the chemical potential,
can only be determined from the explicit evaluation of Eq. (\ref{eq15}).

We now use the Kramers-Kroning dispersion relation for the antifermions,
also imposing a cutoff in the dispersion relation that guarantees
that the momentum of the particle is not far away from the Fermi momentum.
We thus find
\begin{equation}
\label{eq25}
{\rm Re}  \Sigma_- (-E, p) \sim {\rm Re} 
 \Sigma_- (-\mu, p) + \left(\frac{e^2 v^2_F}{12 \pi^2} + 
\frac{e^2 }{32 \pi} \right) \left(\mu -E \right) +
 {\cal O}((\mu -E)^2) \ .
\end{equation}

The leading logarithmic behavior of the real part of the one-loop self-energy of
a quark in the high baryonic limit of QCD  has been obtained
in the ultrarelativistic limit in Ref. \cite{Brown:2000eh}.
There the same leading logarithmic dependence in
the energy as in Eq. (\ref{eq24}) has been found. We should stress here
that this can only be valid for the quark excitations, but not for
the antiquarks ones.

\subsection{Non-Relativistic Limit}

In this subsection we take the non-relativistic (nr) limit of the expressions
computed previously,
restoring  the fundamental constant $c$ in the equations.
The nr limit corresponds to $v_F \ll c$. The antiparticles then decouple.
In such a case, the contribution from the magnetic sector to
the fermion self-energy is suppressed by a factor $(v_F /c)^2$ with respect to the electric sector one. The electric effects are thus dominant.  

The lifetime of an electron $\tau$ is defined as $1/2 \gamma_+$.
We neglect the magnetic contribution to the damping rate, and
express the electric contribution 
in terms of the plasma frequency $\omega_p^2 = \frac 13 M^2 v^2_F$
\begin{equation}
\frac{1}{\tau} = \frac{ \sqrt{3} \pi^2 \omega_p}{32} \left( \frac{E-\mu}{\mu}\right)^2 \frac{c^4}{v^4_F} \ .
\end{equation} 

The relativistic and non-relativistic chemical potentials differ by
the rest mass of the particle, $\mu^2 = \mu^2_{nr} + m^2 c^4$.
For $p^2 \ll m^2 c^2$, $E = mc^2 + \epsilon_{nr} +
 {\cal O} ( \frac{p^4}{m^4 c^2})$, where $\epsilon_{nr} = \frac{p^2}{2m}$.
 Therefore
\begin{equation}
\frac{1}{\tau} \rightarrow
 \frac{ \sqrt{3} \pi^2 \omega_p}{32} \frac{\left(\epsilon_{nr}-\mu_{nr}\right)^2}{m^2 c^4}
 \frac{c^4}{v^4_F} = \frac{ \sqrt{3} \pi^2 \omega_p}{128} \left(
\frac{\epsilon_{nr}-\mu_{nr}}{\mu_{nr}}
\right)^2 \ ,
\end{equation}
where in the last equality we have used $\mu_{nr} = \epsilon_F = \frac 12 m v^2_F$.
The above expression agrees with the computation of the lifetime of
an electron in a non relativistic quantum liquid, using the random
phase approximation (see Eq. (5.134c) of \cite{Pines}).
 
Since in the nr limit 
the difference $(E-\mu) \rightarrow \left(\epsilon_{nr}-\mu_{nr}\right)
+{\cal O} ( \frac{p^4}{m^4 c^2})$,
we also reproduce  the 
dispersion relations due to magnetic interactions of
non relativistic electrons computed in \cite{Holstein}.

\section{Conclusions}

We have derived the one-loop dispersion relations of quasiparticles
and antiquasiparticles
with momentum close to the Fermi momentum in a relativistic electromagnetic
plasma, recovering in the non-relativistic limit the results of Refs.
\cite{Pines,Holstein}. As already emphasized in those papers,
the long-ranged character of the magnetic interactions spoils the
normal Fermi liquid behavior of the plasma. This effect is fully
dominant when the Fermi velocity $v_F$ is close to the velocity of
light. We have also found that the medium modifies in a
different way the propagation properties of particles and
antiparticles. This can be simply
understood from their different phase-space restrictions when they
scatter with the electrons of the Fermi sea.
We should also emphasize that our results are gauge independent.
This is because we have computed the one-loop self-energy on
mass-shell. Off-shell,
the one-loop self-energy (\ref{eq15}) is a gauge-dependent function.

While we have concentrated our study to  QED plasmas, our results
can be easily transported to QCD, only by replacing the electromagnetic
coupling constant by the QCD one and
 taking into account some
additional color factors. In the superconducting phase of QCD, 
the dispersion relations of quarks and antiquarks would be modified
in a different way according to  whether or not these form
Cooper pairs. The dispersion relation for  antiquarks
in the presence of a color  gap has not yet been determined, while
it is still necessary to understand how antiquarks propagate in a
color superconducting medium.

\vskip 2cm

{\bf Acknowledgements:} I want to thank M. Le Bellac, with whom
this work was initiated,  M. Thoma and M. Tytgat for useful discussions.

\appendix

\section{Spectral functions for HDL photon propagators}

The spectral
functions $\rho_{L,T}$ for the resummed propagators $\Delta_{L,T}$
can be found in \cite{LeBellac} for the case
of an ultrarelativistic ($m=0$) plasma. In the case of a
 ultradegenerate plasma,
they can also be determined when $m \neq 0$ \cite{Manuel:1996td}.
For completeness, we will present them below.
In this case, the Debye mass is $M^2 = e^2 \mu^2 v_F /\pi^2$, 
where $v_F$ is the Fermi velocity, defined as the ratio between
the Fermi momentum and the Fermi energy, $v_F = p_F/\mu$. 
The spectral functions of the resummed HDL propagators are computed
from their imaginary part
\begin{equation}
\rho_{L,T} (q_0, {\bf q}) = 2\, {\rm Im} \, \Delta_{L,T} (q_0 + i \epsilon, {\bf q}) \ .
\end{equation}
These functions can be written in terms of a contribution of the poles of
the propagators, plus another one arising from Landau damping:
\begin{equation}
\frac{\rho_{L,T} (q_0,q)}{2 \pi} = Z_{L,T} 
\left[\delta(q_0 - \omega_{L,T} (q)) -
\delta(q_0 + \omega_{L,T} (q)) \right] + \beta_{L,T} (q_0,q) \ .
\end{equation}
The poles $\omega_{L,T}$  are  solutions of the dispersion relations
\begin{mathletters}
\begin{eqnarray}
\omega_L^2(q) & =& \omega_p^2 \frac{3\, \omega_T^2(q)  }{ v^2_F q^2}
\left[  \frac{\omega_T(q)}{2 v_F q}
\ln{\frac{\omega_T(q) + v_F q}{\omega_T(q) - v_F q}} -1
\right] \ ,  \ 0 \leq q < q_{max} \ ,\\
\omega_T^2(q) &=& q^2 + \omega_p^2 \frac{3 \,\omega_T^2(q) }{2 v^2_F q^2}
\left[ 1 +\frac 12 \left( \frac{v_F q}{\omega_T(q)} - \frac{\omega_T(q)}{v_F q}
\right) \ln{\frac{\omega_T(q) + v_F q}{\omega_T(q) - v_F q} }\right]  \ , \ 0 \leq q < \infty \ ,
\end{eqnarray}
\end{mathletters}
where $\omega^2_p= \frac 13 M^2 v^2_F$ is the plasma frequency, and
\begin{equation}
q_{max} = \left( \frac{1}{2 v_F} \ln{\frac{1+ v_F}{1- v_F}} -1
\right)^{\frac{1}{2}} M \ ,
\end{equation}
in the maximum momentum at which the plasmon can propagate.
The above dispersion relations have to be solved numerically. It is
possible to obtain analytically the small $q$ behavior of their
solutions. For $q \ll \omega_p$ 
\begin{equation}
\omega^2_T(q) \rightarrow \omega^2_p + q^2 \left(1 + \frac{v^2_F}{5}
\right) \ , \qquad \omega^2_L(q) \rightarrow \omega^2_p +  \frac{3}{5}
v^2_F q^2 \ .
\end{equation}

The functions $Z_{L,T}$ are the residues of $\Delta_{L,T}$ 
evaluated at their poles, and are given by
\begin{mathletters}
\begin{eqnarray}
Z_L(q) & =& \frac{\omega_L \left( \omega^2_L - v^2_F q^2\right)}{
q^2 \left(3 \omega_p^2 - \left( \omega^2_L - v^2_F q^2\right) \right)} \ , \\
Z_T(q) & =& \frac{\omega_T \left( \omega^2_T - v^2_F q^2\right)}{
3 \omega_p^2 \omega^2_T + \left(\omega^2_T +q^2 \right)
\left(\omega^2_T - v^2_F q^2 \right) - 2\omega_T^2 \left(\omega^2_T -q^2 \right)} \ .
\end{eqnarray}
\end{mathletters}

The pole contribution to the spectral functions is only non-vanishing
above the light-cone.  The Landau damping pieces of the spectral functions
 are non-vanishing only
for $q_0^2 \leq q^2 v^2_F$ and are given by
\begin{mathletters}
\label{betas}
\begin{eqnarray}
\beta_L(q_0,q) & =&
 \frac{M^2\,  x \,\Theta(1-x^2)}{2
\left[ q^2 +M^2  \left( 1 - \frac{x}{2} 
\ln {\Big| \frac{x+1}{x-1} \Big|} \right) \right]^2 + \frac{M^4  \pi^2 x^2 }{4}
} \ , \\
\beta_T(q_0,q) & =&
 \frac{M^2\, v_F^2\, x \,(1-x^2)\Theta(1-x^2)}{
\left[2 q^2(x^2 v_F^2 -1) -M^2 x^2 v_F^2 \left( 1 + \frac{(1-x^2)}{2x} 
\ln {\Big| \frac{x+1}{x-1} \Big|} \right) \right]^2 + \frac{M^4 v_F^4 \pi^2 x^2 (1- x^2)^2}{4}} \ ,
\end{eqnarray}
\end{mathletters}
where $x =q_0/ q v_F$.

\section{Luttinger's Theorem}

The dependence on $(E-\mu)^2$ of the damping rate of a fermion with
energy above $\mu$ can be understood completely as
arising from phase-space restrictions of fermion-fermion scattering,
in the case where the interactions are short-ranged and repulsive.
 The argument,  due to Luttinger \cite{Luttinger}, is simple. 
 We present it below. Let us 
consider the decay rate of a fermion with energy $E$ which interacts
with a fermion with energy $E_k$ inside the Fermi sea. As a result,
two new particles appear, with energies $E_{k'}$ and $E_{p'}$, 
respectively,  which 
are outside the Fermi sea. The total decay rate is then given by
\begin{eqnarray}
\Gamma (E)& = & \frac{1}{E}  \int{{\rm d}^3p'\over (2\pi)^3}
\frac{ \left( 1- \Theta(\mu -E_{p'}) \right)}{2 E_{p'}}
\int{{\rm d}^3 k \over (2\pi)^3} 
\frac{  \Theta(\mu -E_{k})}{2 E_{k}}
\int{{\rm d}^3 k' \over (2\pi)^3} 
\frac{ \left( 1- \Theta(\mu -E_{k'}) \right)}{2 E_{k'}} \\
& \times& (2 \pi)^4 \delta^{(4)} (P +K-P'-K') |{\cal M}|^2 \ ,
\nonumber
\end{eqnarray}
where $|{\cal M}|^2$ is the scattering matrix element squared.
After performing the $p'$ integral
\begin{eqnarray}
\Gamma (E)& = & \frac{2 \pi}{E}  
\int{{\rm d}^3 k \over (2\pi)^3} 
\frac{  \Theta(\mu -E_{k})}{2 E_{k}}
\int{{\rm d}^3 k' \over (2\pi)^3} 
\frac{ \left( 1- \Theta(\mu -E_{k'}) \right)}{2 E_{k'}}
\frac{ \left( 1- \Theta(\mu -E_{{\bf p}+{\bf k}-{\bf k'}}) \right)}{2 E_{
{\bf p}+{\bf k}-{\bf k'}}}
 \\
& \times& \delta (E+E_k-E_{{\bf p}+{\bf k}-{\bf k'}}-E_{k'}) |{\cal M}|^2 \ .
\nonumber
\end{eqnarray}
We now make the change of variables
\begin{equation}
E_k = \mu- t_k \ , \qquad E_{k'} = \mu + t_{k'} \ , \qquad
E_{{\bf p}+{\bf k}-{\bf k'}}= \mu + t_{
{\bf p}+{\bf k}-{\bf k'}} \ ,
\end{equation}
where the $t_i$ variables are positive quantities. The delta function
of energy conservation imposes
\begin{equation}
E-\mu = t_k + t_{k'} + t_{
{\bf p}+{\bf k}-{\bf k'}} \ ,
\end{equation}
which is only valid for $E - \mu \geq 0$. The maximum value that each
one of the variables $t_i$ can achieve is $E-\mu$, while the minimum is zero.
 Using the energies of the particles
as integration variables, we see that the integration
 is always performed
over an energy shell of thickness $E-\mu$. If $E-\mu \ll \mu$, then
the values of the energy variables inside the integral can be substituted
by the Fermi energy. One then finally reaches 
\begin{equation}
\Gamma(E) \propto \int^{\mu}_{\mu -(E-\mu)} dE_k
 \int_{\mu}^{\mu +(E-\mu)} dE_{k'}   \ ,
\end{equation}
and thus  $\Gamma (E) \propto (E-\mu)^2$.
A similar argument can be applied for a quasihole to get the energy
dependence of its damping rate.

In the case we studied in this article, the electric interactions
can be considered as  short-ranged,
because of Debye screening;   they thus give a 
contribution to the damping rate of electrons as expected from 
Luttinger's theorem. The above arguments fail in the case of
magnetic interactions, as those  are not short-ranged, but rather
suffer a weak dynamical screening, where the energies themselves
play the role of infrared cutoffs in  the above integrals.

\end{document}